\documentclass[prl,twocolumn,showpacs,superscriptaddress,amsmath,amssymb,floatfix]{revtex4}
\usepackage{graphicx}
\usepackage{bm}
\begin{document}
\title{Low-energy
quasiparticle excitations in dirty d-wave superconductors and the
Bogoliubov-de Gennes kicked rotator}
\author{\.{I}. Adagideli}
\affiliation{Instituut-Lorentz, Universiteit Leiden, P.O. Box 9506, 2300 RA
Leiden, The Netherlands}
\author{Ph. Jacquod}
\affiliation{D\'epartement de Physique Th\'eorique,
Universit\'e de Gen\`eve, CH-1211 Gen\`eve 4, Switzerland}
\date{\today}
\begin{abstract}
We investigate the quasiparticle density of states in disordered d-wave
superconductors. By constructing a quantum map describing the
quasiparticle dynamics in such a medium, we explore deviations
of the density of states from its universal form ($\propto E$),
and show that additional low-energy quasiparticle states exist provided
(i)~the range of the impurity potential is much larger than the Fermi
wavelength [allowing to use recently developed semiclassical methods];
(ii)~classical trajectories exist along which the pair-potential
changes sign; and (iii)~the diffractive scattering length
is longer than the superconducting coherence length. In the
classically chaotic regime, universal
random matrix theory behavior is restored by quantum
dynamical diffraction which shifts
the low energy states away from zero energy, and the quasiparticle
density of states exhibits a linear pseudogap below an
energy threshold $E^* \ll \Delta_0$.
\end{abstract}
\pacs{71.23.-k, 74.72.-h, 05.45.Mt \\[-0.7cm]}
\maketitle

In recent years, considerable attention has been focused on the
low-energy properties of the quasiparticle spectrum of  disordered cuprate
superconductors~\cite{Hirsch02,Altland02}.
Because many of the cuprate superconductors are randomly
chemically doped insulators and disorder
is a pair-breaker for d-wave superconductors, the role of
nonmagnetic impurities is particularly
important for an understanding of the d-wave superconducting state, its
quasiparticle spectral and transport properties.
Of special interest is the low-energy behavior of the
single-particle Density of States (DoS) $\rho(E)$.

In early work~\cite{REF:nerseyan}, the self consistent $T$-matrix
approximation was shown to break down for 2-dimensional d-wave
superconductors. This led to a series of papers using
nonperturbative methods, which predicted (at first sight)
contradictory results:
vanishing~\cite{REF:nerseyan,REF:Senthil,REF:Bocquet,REF:Altland},
constant~\cite{REF:Ziegler98,REF:Huckenstein}, and
diverging~\cite{REF:Pepin,Ada02} DoS as $E\rightarrow 0$.
On the numerical side, several investigations also predicted both
vanishing~\cite{Ghos00,Atkinson01}, and
diverging~\cite{REF:detofdisord,REF:zhu} DoS at zero energy. It
was soon argued, based on numerical analysis, that the reason
behind these contradicting predictions is the fact that the
microscopic details of disorder (i.e. details beyond the
transport mean free path $\ell$, such as the density of scatterers
or the correlation length $\zeta$ of the impurity potential) as
well as the symmetries of the clean Hamiltonian matter both
qualitatively and quantitatively~\cite{REF:detofdisord,REF:Yashenkin01} (see
also~\cite{Chamon01}). An important feature shared by the numerics
of Refs.~\cite{Ghos00,Atkinson01,REF:detofdisord,REF:zhu,REF:Yashenkin01}
is that the disorder is introduced via isolated {\it point-like\/}
scatterers. 
Long-wavelength disorder, which may arise due to chemical doping
away from ${\rm CuO}_2$ planes, or can be
induced via ion radiation
techniques~\cite{REF:ionrad} or via a STM tip~\cite{REF:Yazdani}, is thus ignored. 
Effects of long-wavelength disorder are
expected to become dominant when the ${\rm CuO}_2$ planes
have almost no atomic disorder~\cite{REF:clean};
they are the main focus of this paper.

It has recently been realized, in the context of mesoscopic
physics and weak localization, that $\zeta$ and $\ell$, together
with the Fermi wavelength $\lambda_F$, define
two classes of complex quantum systems: quantum
disordered systems where $\lambda_F \ell/\zeta^2 > 1$ and
quantum chaotic systems for which $\lambda_F \ell/\zeta^2 < 1$~\cite{Alarkin}.
The latter class is characterized by the emergence of a new
{\it diffractive scattering} time
scale, $\tau_E=\nu^{-1} \ln [\zeta/\lambda_F]$,
defined as the time it takes for the classically chaotic dynamics
(with Lyapunov exponent $\nu$) to stretch a wavepacket
of minimal initial extension $\lambda_F$ to a length $\zeta$.
In contrast to quantum disordered systems, quantum chaotic systems exhibit
nonuniversal properties due to their short-time classical (i.e. deterministic)
dynamics.
In particular, significant
deviations from Random Matrix Theory (RMT) emerge, as was
recently found by Adagideli et al.~\cite{Ada02} in the context of
impurities in d-wave superconductors. These authors
used a semiclassical approach to calculate the
low energy DoS for a collection of {\it extended\/} scatterers (a quantum
chaotic system) and found an asymptotic behavior
$\rho(E)\sim 1/E|\ln E^2|^3$ as $E \rightarrow 0$.
They nevertheless argued that the RMT predictions of a linear
pseudogap would be restored at lower energy $E<E^*$, i.e. that
the singularity in the DoS
would be cut off at an energy $E^*$ related to $\tau_E$,
by diffractive (nonclassical) scattering occurring at larger
times $\tau > \tau_E$.
The purpose of the present paper is to investigate the modifications that
the DoS undergoes as the correlation length of the impurity
potential increases and $\tau_E$ becomes relevant. We will
focus our attention on (i) providing for numerical checks of the
theory of Adagideli et al. in the case of long-wavelength
disorder~\cite{Ada02}; (ii) finding out whether
for some $E^*$ the DoS
is suppressed for $E<E^*$, in agreement with RMT predictions
\cite{REF:Senthil,REF:Bocquet,REF:Altland}, (this could also reconcile the
above mentioned contradictory predictions);
and (iii)~investigate the transition region between extended and
pointlike disorder.

We start by introducing a quantum map model for
quasiparticle states in disordered d-wave superconductors.
The main motivation behind this model is to investigate
discrepancies (in low-energy DoS) between point-like vs. extended
disorder as well as the
transition region, i.e.~the regime in which the impurity size is
intermediate. To the best of
our knowledge no disorder model which includes extended impurities
as well as pointlike
in d-wave superconductors has been studied numerically so far.
This map has two additional advantages:
First, as both the density and correlation length
of impurities can be tuned independently, it is possible to interpolate
between the two
extreme regimes of strong disorder: unitary disorder (i.e.~disorder due
to dilute, pointlike
scatterers~\cite{REF:Pepin,Chamon01,REF:Yashenkin01,REF:Altland}, {\it viz.}~quantum
disorder)
and quasiclassical disorder~\cite{Ada02} (i.e.~disorder due to
extended scatterers, {\it viz.}~quantum chaotic). Second, from a numerical point of view, it
allows for the investigation of very large system sizes, i.e. lattice
sizes of up to $256\times 256$, which are necessary for both variations
of the disorder correlation length and the numerical extraction
of the parametric behavior of the DoS.
Our reasons why a dynamical model is relevant are:
(i)~In absence of superconductivity, many properties
of quasiparticles in disordered media (such as Anderson localization)
are correctly described by 1-D maps~\cite{Fish82}. In fact it has been shown
by Altland and Zirnbauer that one of those maps, the 1-D kicked rotator,
and quasi-one-dimensional metallic wires are described by the
same effective field theory~\cite{Alt96}.
~Recent numerical investigations in
$D=2$ suggest that this is also true
in higher dimensions~\cite{Ossi02}.
(ii)~In presence of superconductivity,
Andreev maps based on the kicked rotator
have recently been shown to adequately describe quantum dots in
contact with a superconductor~\cite{Jac03}.

We first briefly discuss generic properties of quantum maps
for uncoupled quasiparticles. The dynamics
corresponds to a succession of free propagations,
interrupted by sudden {\it kicks} of period $\tau_{0}$,
i.e. instantaneous perturbations.
Quantum maps are conveniently represented by a unitary, {\it Floquet}
operator $F$, giving the time-evolution after $p$ kicks
as $u(p)=F^{p}u(0)$, for an initial wavefunction
$u(0)$. The matrix $F$ has eigenvalues $\exp(-i\varepsilon_{m})$, which define
quasi-energies $\varepsilon_{m}\in (-\pi,\pi)$ 
(energies and quasi-energies are expressed in units of $\hbar/\tau_{0}$).
While the energy is not conserved, the periodicity of the kick
still preserves quasi-energies, much in the same way as a periodic
potential breaks translational symmetry, but still preserves
quasi-momentum. Time evolution of hole excitations (being the time-reversed of electronic
excitations) is given by $v(p)=(F^{\ast})^{p}v(0)$.
Specializing to the $D$-dimensional
kicked rotator, we write the Floquet operator
as \cite{Izr90}
\begin{eqnarray}
F&=&
\exp\left(-i\; \frac{K I}{\hbar \tau_0} \; \Pi_{j=1}^D\cos r_j \right)
\exp\left(i\frac{\hbar\tau_{0}}{2 I} \vec{\nabla}^2 \right).
\label{kickedF}
\end{eqnarray}
It describes the free motion of a particle with
dimensionless coordinates $\{r_j\}$ (e.g. expressed in units of
a lattice constant), which is
interrupted at periodic time intervals $\tau_{0}$ by a kick
of strength $K \cdot \Pi_{j=1}^D \cos r_j$. $I$ is the moment of inertia
of the particle,
and $K$ is the kicking strength.
For $D=1$ and 2, increasing $K$ makes the classical dynamics evolve from 
integrable ($K=0$) to fully chaotic [$K\agt 7$, with Lyapunov exponent
$\lambda\approx\ln (K/2)$].
For $0<K<7$ stable and unstable motion coexist
(a so-called mixed phase space)~\cite{Izr90}. Increasing $K$ is thus
tantamount to increasing the amount of disorder, the fully chaotic
regime corresponding to a finite density of impurities.

Electron and hole excitations inside a superconductor
are however coupled by a nonvanishing pair-potential.
Accordingly we extend the kicked rotator of Eq.\ (\ref{kickedF})
to a Bogoliubov-de Gennes form. We discuss this
construction for the case $D=2$.
First, we replace the free quasiparticle motion by a coupled electron
and hole dynamics,
\begin{subequations}
\label{eq:bdgfree}
\begin{eqnarray}
{\cal F}_0&=&\exp(-i {\cal H} \tau_{0}/\hbar), \\
{\cal H}&=& H \; \sigma_z + \Delta \; \sigma_x.
\end{eqnarray}
\end{subequations}
Here, $H = -(\hbar^2 \vec{\nabla}^2/2 I)-E_F$, with
$E_F$ the Fermi energy,
$\sigma_{x,z}$ are Pauli
matrices acting in particle-hole space, and $\Delta$ is the
superconducting pair potential.
Second, the coupled quasiparticle motion is followed by a kick
\begin{subequations}
\label{kicksemicl}
\begin{eqnarray}
{\cal F}_K&=&\exp(-i {\cal H}_K/\hbar), \\
{\cal H}_K&=& \frac{K I}{\tau_0}
\cos x \cos y  \; \sigma_z.
\end{eqnarray}
\end{subequations}
Exponentiating the Pauli matrices, we end up with the
Bogoliubov-de Gennes-Floquet (BdGF) operator
\begin{subequations}
\label{eq:bdgcoupled}
\begin{eqnarray}
{\cal F} &=&
{\cal F}_K \; {\cal F}_0 ,\\
{\cal F}_0 &=& \cos \sqrt{(H^2+\Delta^2)(\tau_0/\hbar)^2 } \;\; {\cal I}
\nonumber \\
& + & \frac{i \sin\sqrt{(H^2+\Delta^2)(\tau_0/\hbar)^2 }}{\sqrt{H^2+\Delta^2}} \; [
H \sigma_z + \Delta \sigma_x ], \\
{\cal F}_K&=& \cos\left[\frac{K I}{\hbar \tau_0} \cos x \cos y \right]
\; {\cal I} \nonumber \\
& + &  i
\sin\left[\frac{K I}{\hbar \tau_0} \cos x \cos y \right] \; \sigma_z,
\end{eqnarray}
\end{subequations}
with ${\cal I}$, the identity matrix in particle-hole space.
For $\Delta=0$, Eq.\ (\ref{eq:bdgcoupled})
describes uncoupled
electron and hole excitations in a disordered 2$D$ metal.
Once this metal becomes
superconducting, $\Delta$ couples these excitations
during their free propagation, while it is neglected during
the instantaneous kick. As in the case of a BdG eigenproblem,
the $2M$ quasienergies [with
average spacing $\delta\equiv \langle \varepsilon_{m+1}-\varepsilon_m \rangle
=\pi/M$] of the BdGF equation
${\cal F} \phi_m = \exp(-i \varepsilon_m) \phi_m$, come in pairs
with opposite sign $\varepsilon_m=-\varepsilon_{2M-m+1}$, similarly
to the spectral properties of a BdG Hamiltonian. These considerations
establish the correspondence between the map of Eq.\ (\ref{eq:bdgcoupled})
and quasiparticles in a dirty superconductor.

\begin{figure}
\includegraphics[width=8cm]{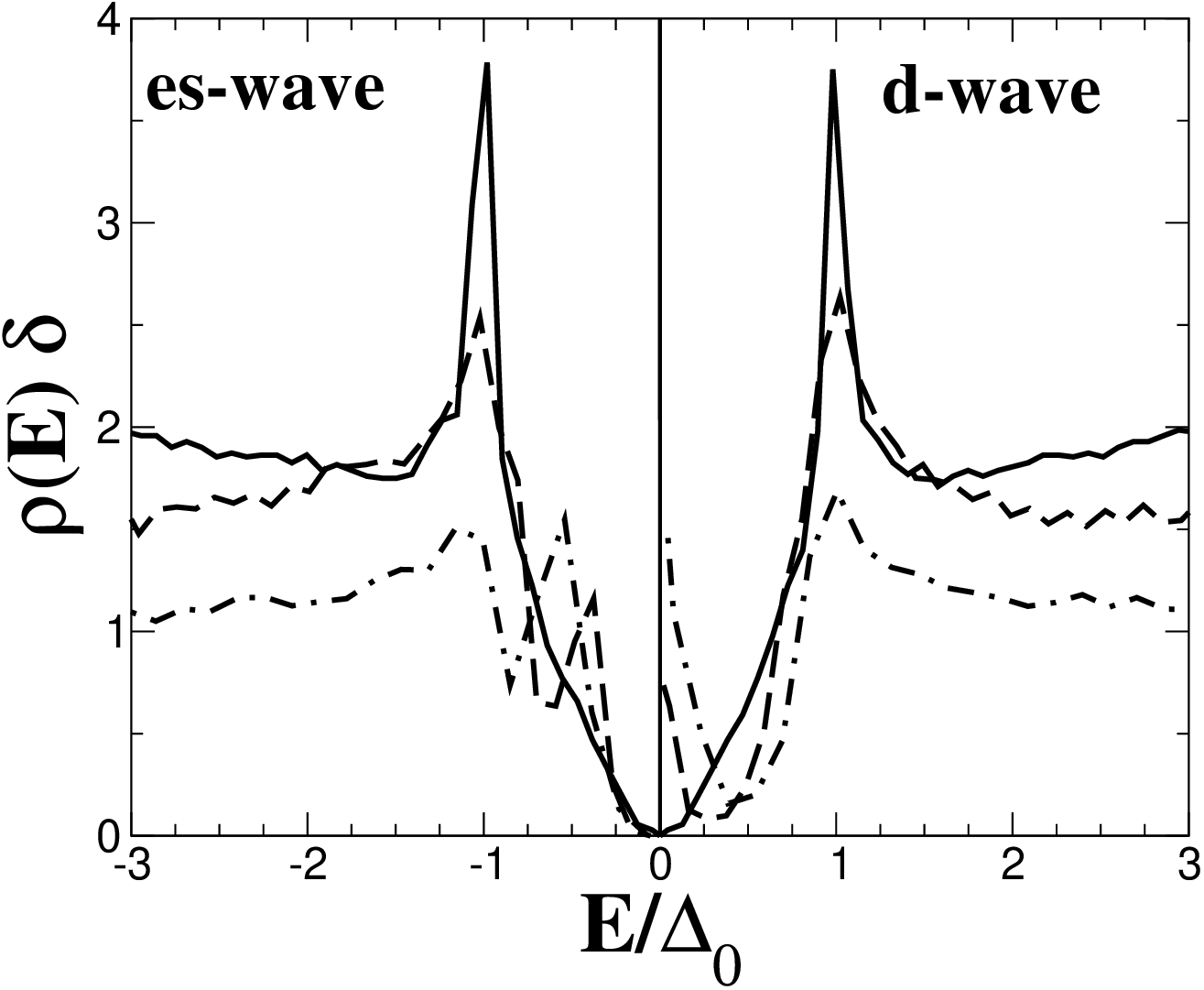}
\caption{Density of states for the d-wave (right) and
extended s-wave (left) Bogoliubov-de Gennes kicked rotator
defined by
Eqs.\ (\ref{eq:bdgfree}-\ref{eq:bdgcoupled}),
and parameters
$L_x \times L_y = 128 \times 128$, $\Delta_0=0.4$, $E_F=2 \pi^2/5$,
and $K=0.$ (solid lines), 2. (dashed lines) and 8. 
(dotted-dashed lines).\\[-0.7cm]}
\label{figure1}
\end{figure}

We next quantize the phase
space on a $4$-torus $\{x,y;p_x,p_y\}$, with dimensionless
momentum
$p_{x,y} =-i\hbar_{\rm eff} \partial/\partial(x,y) \in (0,2 \pi)$~\cite{Izr90}.
The effective Planck
constant $\hbar_{\rm eff}\equiv\hbar\tau_{0}/I_{0}$ takes on values
$\hbar_{\rm eff}=2\pi/M$, with integer $M=L_x \cdot L_y$, in term
of the real-space linear system sizes $L_{x,y}$ (also expressed in units
of a lattice spacing), and the impurities have a spatial extension
$\zeta=O(L_x,L_y)$.
The BdGF operator is then a $2M \times 2M$ unitary
matrix, and we consider the two cases of d-wave
$[\Delta(p_x,p_y)=\Delta_0 (p_x^2-p_y^2)/(p_x^2+p_y^2)]$ and
extended s-wave $[\Delta(p_x,p_y)=\Delta_0 |p_x^2-p_y^2|/(p_x^2+p_y^2)]$
pair potentials, for which ${\cal F}_0$ is diagonal in
momentum representation. Noting that ${\cal F}_K$ is diagonal
in real space representation, we rewrite ${\cal F}$ as
\begin{eqnarray}
\label{nummap}
{\cal F}_{\vec{p}\vec{p}'} &=& (\left[{\cal U}
{\cal F}_K {\cal U}^\dagger \right] \; {\cal F}_0)_{\vec{p}\vec{p}'},
\end{eqnarray}
where ${\cal U} = U {\cal I}$,
and $U$ is the unitary matrix of the 2$D$ Fourier
transform between real space and momentum coordinates,
$U_{\vec{p}\vec{p}'} = M^{-1/2} \exp[(2 \pi i/M)\; \vec{p} \cdot \vec{p}\;']$.
We numerically extract the quasienergy DoS from the eigenvalues
$\sin \varepsilon_m$ of the hermitean matrix
$\frac{1}{2 i}({\cal F}-{\cal F}^{\dagger})$, which we
diagonalize using the Lanczos algorithm \cite{Ket99}.

In Fig.~1 we show the quasienergy DoS for d-wave and extended s-wave
pair potentials away from half filling ($E_F = 2 \pi^2/5 < \pi^2/2$),
as the kicking strength increases. In the clean case
($K=0$) the two DoS are the same. The gap singularity at $E/\Delta_0=1$ gets
washed out as $K$ increases in both cases, however, a peak
emerges in the d-wave DoS around $E=0$, while $\rho(E)=0$ in the
extended s-wave case. This is in agreement with
Ref.~\cite{Ada02,Ada99}, i.e. the existence of low-energy states requires
a change in the sign of the pair potential. In the
extended s-wave case, the low energy peak is shifted
by an energy corresponding to
the gap averaged over all momenta mixed by the impurity potential.

\begin{figure}
\includegraphics[width=8cm]{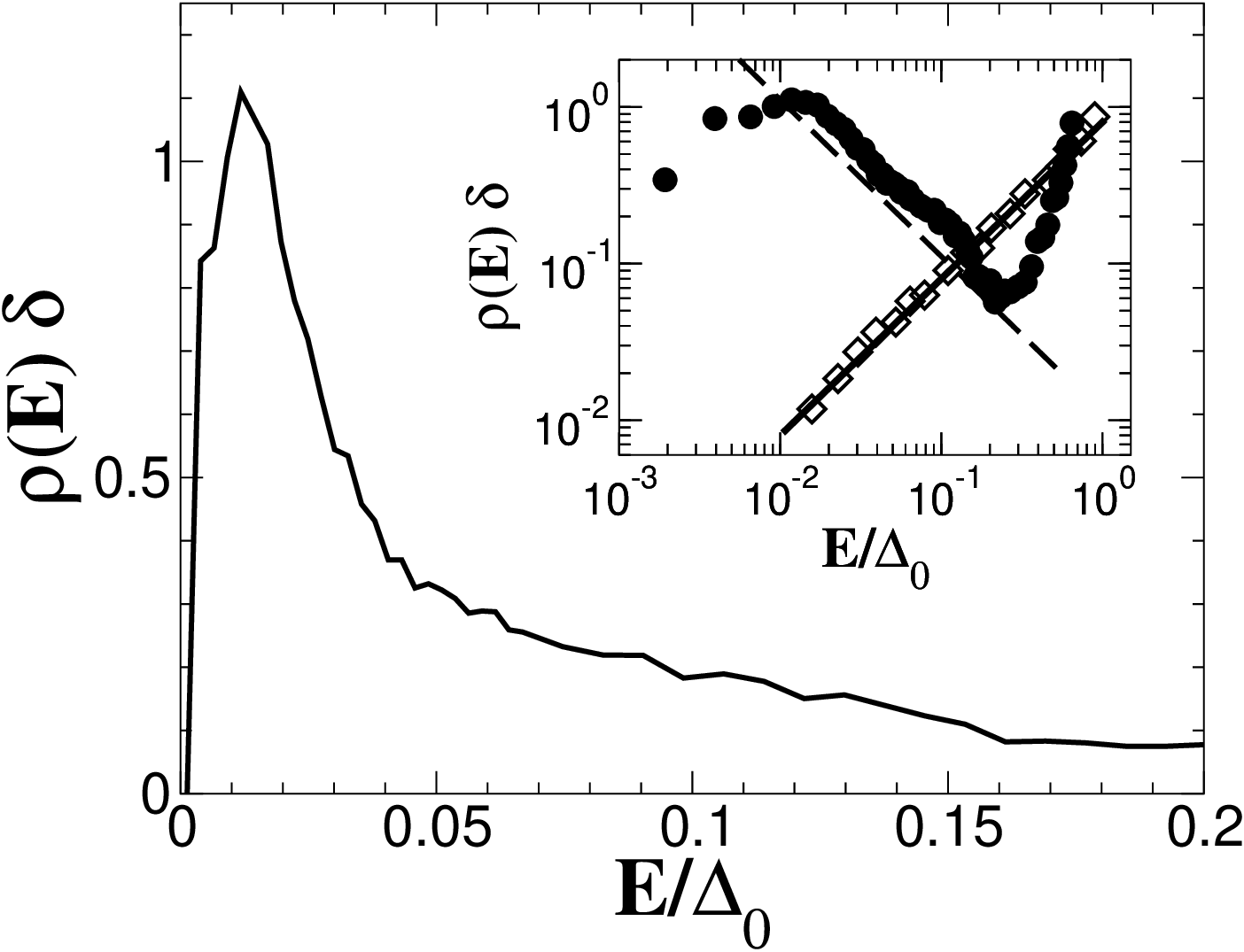}
\caption{Main plot: Low energy DoS for the d-wave
Bogoliubov-de Gennes kicked rotator of
Eqs.\ (\ref{eq:bdgfree}-\ref{eq:bdgcoupled}),
$L_x \times L_y = 256 \times 256$, $\Delta_0=0.4$, $E_F=2 \pi^2/5$,
and $K=8$.
Inset: Asymptotic of the density of states at low excitation
energy for the same set of parameters, for
long-wavelength disorder (quantum chaotic; black circles) and 
diffractive disorder as defined in Eq. (\ref{kickdiffr})
with $N_H=27$ (quantum disorder; empty diamonds).
The solid and dashed lines indicate a $\propto E$ and $\propto E^{-1}$
behavior respectively.\\[-0.8cm]}
\label{figure2}
\end{figure}

We focus on the d-wave symmetry from now on. A closer look 
at the DoS in the fully chaotic regime with $K=8.$
is provided in Fig.~2. It indicates
that the characteristic semiclassical
singularity exhibited by the DoS as $E \rightarrow 0$ is cut off
at an energy $E^* \ll \Delta_0$,
where a sharp drop occurs and $\rho(E) \rightarrow 0$.
RMT predicts such a drop to occur over an energy scale given
by the Thouless energy~\cite{Altland02}, which in our case
is however significantly larger than $\Delta_0$~\cite{caveat}.
We thus attribute this drop to the emergence of
diffractive scattering
at times larger than $\tau_E$~\cite{Alarkin} as follows. According to
Ref.~\cite{Ada02}, the DoS corresponding to
low energy semiclassical states can be estimated from a mapping
onto a tight-binding chain with random hoppings, for which the
eigenfunctions are localized with an energy-dependent
localization length $\xi(E)$ \cite{Eggarter}.
At low energies, $\xi$ exceeds the diffractive scattering length 
$v_F \tau_E$, ($v_F$ is the
Fermi velocity) in which case hoppings between otherwise uncoupled
tight-binding chains (corresponding to different
classical trajectories) have to be taken into account. The emergence of 
these processes signals the breakdown of semiclassics and the restoration of
RMT. One thus expects
the vanishing of the DoS below a threshold energy given by
the condition $\xi(E^*) \approx v_F \tau_E$. Since $\xi(E)$
is bounded by the superconducting coherence length,
$\xi(E) \gtrsim \hbar v_F/\Delta_0$ \cite{Ada02}, 
the observation of the semiclassical
peak in the DoS requires a long enough diffractive scattering
length $v_F \tau_E >  \hbar v_F/\Delta_0$.
While preliminary results
corroborate this argument, 
a detailed investigation of $E^*$ will be presented elsewhere \cite{Jacada03}.

In the inset to Fig.~2 we show the asymptotic behavior of the
DoS on a log-log scale. Once abstraction is made of the
drop in the DoS below $E^*$, the semiclassical data
exhibit a singular behavior slightly below $\propto E^{-1}$ (black circles)
which is in qualitative agreement with the
prediction $\rho(E) \propto E^{-1} |\ln E|^{-3}$  of Ref.~\cite{Ada02}.

Having established the validity of semiclassical predictions in the
quantum chaotic regime, we next decrease the
range of the disorder and enter the quantum disordered regime.
We accomplish this via the inclusion of higher harmonics to the kicking
potential, and replace Eq.\ (\ref{kicksemicl}) by
\begin{equation}
{\cal H}_K = \frac{K I}{N_H^2 \tau_0}
\sum_{l,m=1}^{N_H} \cos[l x]  \cos[m y] \; \sigma_z.
\label{kickdiffr}
\end{equation}
The typical impurity size
decreases as $\zeta \propto N_H^{-1}$. Fig.~3 shows the disappearance of the
low-energy peak in the DoS as $N_H$ increases. For the set of parameter
considered, once $N_H \simeq 27$ is reached, the DoS vanishes at $E=0$.
Note that the resolution used in Fig.~3 does not allow to see the
opening of the RMT gap below $E^*$.
A more precise look at the DoS for $N_H=27$ is provided
on the inset to Fig.~2 (empty diamonds). 
The data clearly indicate the
expected RMT linear suppression of the DoS.

\begin{figure}
\includegraphics[width=8cm]{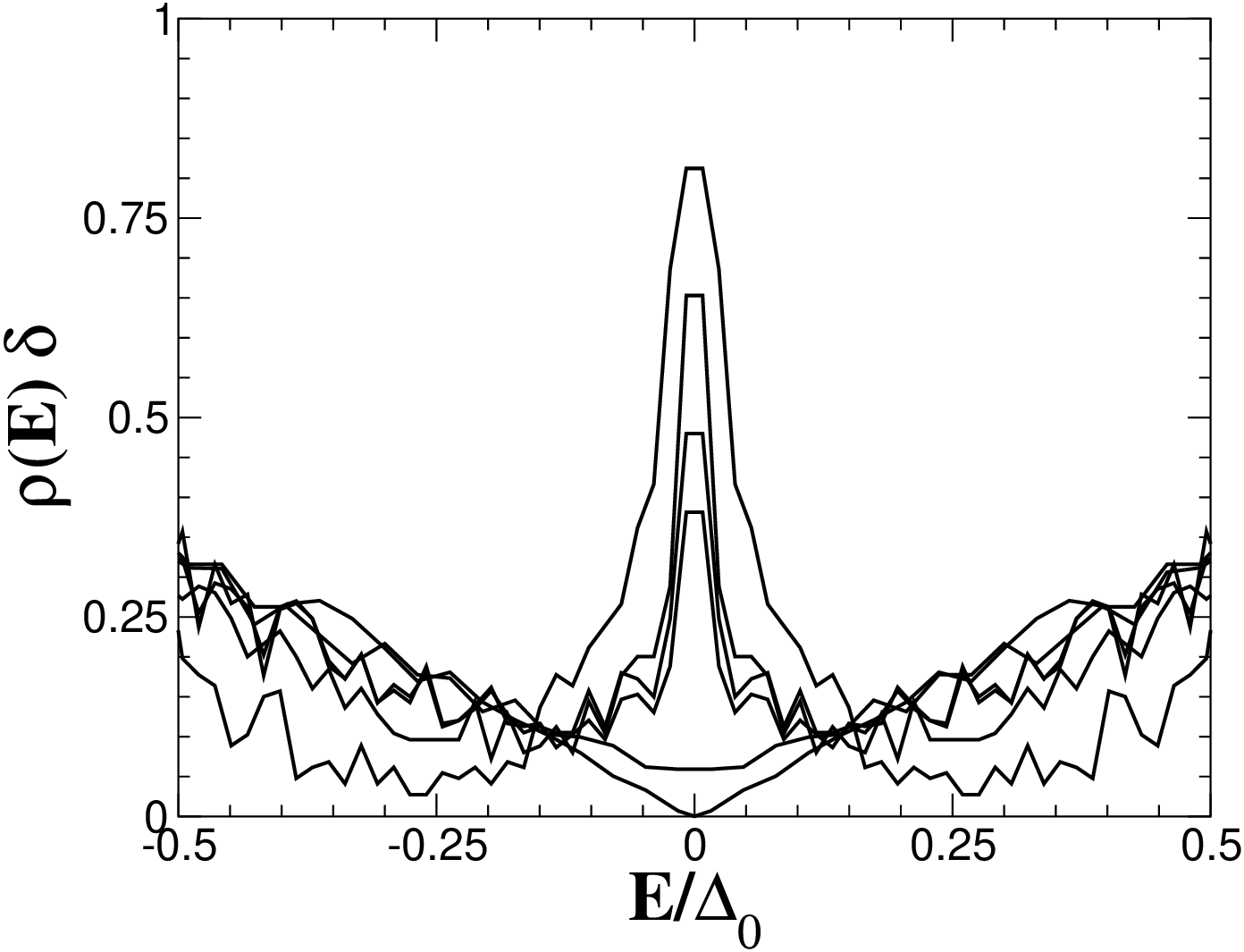}
\caption{Low energy DoS for the d-wave
Bogoliubov-de Gennes kicked rotator with decreasing disorder range as
defined in
Eqs.\ (\ref{eq:bdgfree}), (\ref{eq:bdgcoupled}) and (\ref{kickdiffr}), with
$L_x \times L_y = 256 \times 256$, $\Delta_0=0.4$, $E_F=2 \pi^2/5$,
$K=8.$, and $N_H=1,3,5,7,17$, and 27 (from top to bottom).\\[-0.8cm]}
\label{figure3}
\end{figure}

Our results thus clarify the competition between 
RMT \cite{Altland02} and semiclassics \cite{Ada02}. 
The next step is to investigate
the transport properties and 
to explore the parametric dependence of $E^*$.
Work along those lines is in progress \cite{Jacada03}.

This work was supported by the Dutch Science Foundation NWO/FOM and
the Swiss National Science Foundation. We thank I. Gornyi, 
A. Yashenkin, M. Vojta, N. Trivedi, and P. M. Goldbart for interesting 
discussions and comments.


\begin{thebibliography}{99}
\bibitem{Ket99} This diagonalization is performed
with only $O(M^2 \ln M)$ operations, if the
multiplication with ${\cal U}$ in Eq.\ (\ref{nummap}) is performed with
the Fast-Fourier-Transform algorithm, see:
R. Ketzmerick, K. Kruse, and T. Geisel, Physica D {\bf 131}, 247 (1999).
\bibitem{caveat} Quantizing the kicked rotator on a torus corresponds to a
ballistic chaotic system \cite{Izr90} so that the Thouless
energy is of the order of the size of the band.
\end{thebibliography}
\end{document}